\documentclass[submission,copyright,creativecommons]{eptcs}
\providecommand{\event}{Non-Classical Logics 2024 (NCL'24)}

\usepackage{proof}

\newcommand{\PROOF}{\mbox{\bf Proof.}\ \ }
\newcommand{\QED}{\hfill \rule{1.3ex}{0.6em}}
\newcommand{\SEQ}[2]{\mbox{$ #1 \Rightarrow #2 $}}
\newcommand{\I}{\mbox{$\to$}} 
\newcommand{\IMP}{\mbox{$\to$}} 
\newcommand{\LAND}{\mbox{$\land$}} 
\newcommand{\LOR}{\mbox{$\lor$}}

\newcommand{\BB}{\raisebox{+\height}{\fbox{\phantom{}}}} 
\newcommand{\DIA}{\mbox{$\diamondsuit$}}

\newcommand{\al}{\mbox{\it $\alpha$}}
\newcommand{\be}{\mbox{\it $\beta$}}
\newcommand{\ga}{\mbox{\it $\gamma$}}

\newcommand{\GA}{\mbox{\it $\Gamma$}}
\newcommand{\DE}{\mbox{\it $\Delta$}}
\newcommand{\SI}{\mbox{\it $\Sigma$}}
\newcommand{\PI}{\mbox{\it $\Pi$}}
\newcommand{\LA}{\mbox{\it $\Lambda$}}

\newtheorem{prop}{Proposition}[section] 
\newtheorem{thm}[prop]{Theorem} 
\newtheorem{lm}[prop]{Lemma} 
 
\newtheorem{df}[prop]{Definition} 
\newtheorem{eg}[prop]{Example} 
\newtheorem{rmk}[prop]{Remark}

\usepackage{iftex}

\ifpdf
  \usepackage{underscore}         
  \usepackage[T1]{fontenc}        
\else
  \usepackage{breakurl}           
\fi

\title{Twist Sequent Calculi for S4 and its Neighbors}
\author{Norihiro Kamide
\institute{School of Data Science, Nagoya City University, Aichi, Japan}
\email{drnkamide08@kpd.biglobe.ne.jp}
}
\def\titlerunning{Twist Sequent Calculi for S4 and its Neighbors}
\def\authorrunning{Norihiro Kamide}
\begin{document}
\maketitle

\begin{abstract}
Two Gentzen-style twist sequent calculi for the normal modal logic S4 are introduced and investigated. The proposed calculi, which do not employ the standard logical inference rules for the negation connective, are characterized by several twist logical inference rules for negated logical connectives. Using these calculi, short proofs can be generated for provable negated modal formulas that contain numerous negation connectives. The cut-elimination theorems for the calculi are proved, and the subformula properties for the calculi are also obtained. Additionally, Gentzen-style twist (hyper)sequent calculi for other normal modal logics including S5 are considered. 
\end{abstract}

\section{Introduction}
\label{introduction-section}

Reasoning about negative information or knowledge, especially when involving negations and modalities, holds significant importance in the field of philosophical logic \cite{FITTING-FI-1991,Wansing-JPL-2002,OdintsovWansing2010,RivieccioJungJansana2015,DW-RSL-2020}. For instance, Fitch's paradox, a fundamental issue in philosophical logic, has been analyzed through reasoning about negative information within the context of negations and modalities \cite{Wansing-JPL-2002}. Effective reasoning in this area requires the development of a robust proof system, such as a Gentzen-style sequent calculus, tailored for standard modal logics like the normal modal logic S4. This Gentzen-style sequent calculus should efficiently manage the interactions between negations and modalities.

The primary objective of this study is to develop an alternative cut-free and analytic Gentzen-style sequent calculus for S4. Specifically, the sequent calculus proposed in this study aims to effectively handle negative information involving negations and modalities. In other words, our focus is on constructing a sequent calculus capable of managing formulas that include both modal operators and multiple negation connectives. The proposed sequent calculi are intended to have the ability to generate relatively short and compact ``shortcut (or abbreviated) proofs'' for provable negated modal formulas containing numerous negation connectives.

The concept of a ``shortcut (or abbreviated) proof'' is defined as a proof that incorporates ``twist logical inference rules.'' These twist rules are considered ``shortcut (or abbreviated) rules'' specifically in relation to negations. To explain these twist rules, we now examine the following twist logical inference rule for negated modal operators, which is included in one of the proposed calculi, gTS4: 
$$
\infer[(\neg\BB{\rm left}^{T}).]{ \SEQ{\neg\BB\al, \BB\GA_1, \neg\DIA\GA_2}{\DIA\DE_1, \neg\BB\DE_2}}{
  \SEQ{\BB\GA_1, \BB\DE_2}{\DIA\DE_1, \DIA\GA_2, \al}
}
$$
This rule is derivable in a standard sequent calculus as follows: 
$$
\infer*[(\neg {\rm left}), (\neg {\rm right})]{ \SEQ{\neg\BB\al, \BB\GA_1, \neg\DIA\GA_2}{\DIA\DE_1, \neg\BB\DE_2}}{
      \infer[(\BB {\rm right}^k)]{\SEQ{\BB\GA_1, \BB\DE_2}{\DIA\DE_1, \DIA\GA_2, \BB\al}}{ 
                \SEQ{\BB\GA_1, \BB\DE_2}{\DIA\DE_1, \DIA\GA_2, \al}
      }
}
$$
where ($\neg${\rm left}), ($\neg${\rm right}), and ($\BB${\rm right}$^k$) 
\footnote{
 ($\BB$right$^k$) was originally introduced by Kripke in \cite{KRIPKE-ZML1963} (p. 91). 
}
are defined as follows: 
$$
\infer[(\neg {\rm left})]{\SEQ{\neg\al, \GA}{\DE}}{
    \SEQ{\GA}{\DE, \al}
}
\quad
\infer[(\neg {\rm right})]{\SEQ{\GA}{\DE, \neg\al}}{
    \SEQ{\al,\GA}{\DE}
}
\quad
\infer[(\BB {\rm right}^k).]{\SEQ{\BB\GA}{\DIA\DE, \BB\al.} }{
  \SEQ{\BB\GA}{\DIA\DE, \al}
}
$$

In this case, we can observe that the applications of the rules ($\neg$left), ($\neg$right), and ($\BB$right$^k$) are encapsulated within the single rule ($\neg\BB$left$^T$). Specifically, ($\neg\BB$left$^T$) serves as a shortcut (or abbreviated) rule for the applications of ($\neg$left), ($\neg$right), and ($\BB$right$^k$). In other words, many applications of ($\neg$left) and ($\neg$right) in a proof can be abbreviated by a single application of ($\neg\BB$left$^T$). Therefore, if there are many occurrences of $\neg$ in a given provable sequent, we can obtain a significantly shorter shortcut (or abbreviated) proof for the sequent compared to using the standard calculus. In this sense, gTS4 is effective in proving negated modal formulas containing numerous negation connectives.

In this study, we introduce two cut-free and analytic Gentzen-style twist sequent calculi for the modal logic S4, named lTS4 and gTS4. These calculi handle negation differently: locally in lTS4 and globally in gTS4. Both lTS4 and gTS4 avoid using standard logical inference rules for negation. Instead, they incorporate several twist logical inference rules, which serve as shortcut (or abbreviated) rules specifically designed for handling negated logical connectives. These twist rules are constructed by integrating the standard logical inference rules for the logical connectives $\LAND, \LOR, \IMP, \neg$ and the modal operators $\BB, \DIA$ with those for $\neg$.

Due to these twist logical inference rules, lTS4 and gTS4 can generate relatively short and compact shortcut (or abbreviated) proofs for provable negated modal formulas containing multiple negation connectives. This makes lTS4 and gTS4 particularly effective in handling negated modal formulas. Indeed, the proofs produced by lTS4 and gTS4 for the sequents that include negated modal formulas containing numerous negation connective are shorter than those generated by a standard Gentzen-style sequent calculus for S4. Thus, we can understand that lTS4 and gTS4 have the ability to provide effective (shortcut or abbreviated) reasoning in this context.

In this study, we establish the cut-elimination theorems for both lTS4 and gTS4, confirming that they are cut-free. Additionally, we demonstrate the subformula properties for these calculi, ensuring that lTS4 and gTS4 are analytic. Furthermore, we extend similar results to some Gentzen-style twist sequent calculi designed for classical logic and other normal modal logics, including K, KT, and S5. Specifically, a Gentzen-style twist sequent calculus for classical logic, called TCL, is obtained as the common fragment of lTS4 and gTS4 when the modal operators $\BB$ and $\DIA$ are omitted.

We now examine some closely related traditional and recently proposed Gentzen-style sequent calculi for S4. A cut-free and analytic Gentzen-style sequent calculus for S4 was initially introduced and investigated by Ohnishi and Matsumoto in \cite{OM1957,OM1959}. Another cut-free and analytic Gentzen-style sequent calculus, referred to here as GS4, was presented by Kripke in \cite{KRIPKE-ZML1963} (p. 91). Kripke's calculus GS4 was developed by adapting Ohnishi and Matsumoto's calculus to handle the modal operators $\BB$ and $\DIA$ simultaneously. Grigoriev and Petrukhin introduced and explored some extensions of GS4 in \cite{GP-LLP2019}, wherein some multilattice extensions of GS4 and its S5 version were studied.

Cut-free (though non-analytic) Gentzen-style sequent calculi NS4, DS4, and SS4 for S4, which are regarded as falsification-aware calculi, have been introduced by Kamide in \cite{KAMIDE-JLLI-2022}, based on GS4. Furthermore, cut-free (though non-analytic) Gentzen-style sequent calculi GS4$_1$, GS4$_2$, and GS4$_3$ for S4, which are compatible with a Gentzen-style sequent calculus for Avron's self-extensional paradefinite logic, have also recently been introduced by Kamide in \cite{Kamide-JAL-2024}, based on GS4.

The original calculi introduced by Ohnishi and Matsumoto and by Kripke were cut-free and analytic systems, yet they were not effective in proving negated modal formulas containing numerous negation connectives. While NS4, DS4, and SS4 were suitable for falsification-aware reasoning and GS4$_1$, GS4$_2$, and GS4$_3$ were compatible with paraconsistent reasoning, they were not effective for proving negated modal formulas containing numerous negation connectives. Moreover, NS4, DS4, GS4$_1$, GS4$_2$, and GS4$_3$ lacked analyticity (i.e., these calculi lacked the subformula property).

In contrast to these calculi, the proposed twist calculi, lTS4 and gTS4, are cut-free, analytic, and effective in proving negated modal formulas containing numerous negation connectives. For more general information on sequent calculi for modal logics including S4, see, for example, \cite{Wansing-2002,Fitting-2007,Poggiolesi-2010,Negri-2011,I-2019,MMZ-ACM-2021,MMZ-BSL-2023,LP-2024} and the references therein. For information on sequent calculi for S5, see, for example, \cite{GP-LLP2019,KAMIDE-JLLI-2022,Poggiolesi-2010,LP-2024,MMZ-ACM-2021,MMZ-BSL-2023,I-2019} and the references therein. For a very short survey of recent works on sequent calculi for S5, see Section \ref{conclusion-lTS4} of the present paper.

The structure of this paper is addressed as follows. 

In Section \ref{twist-calculus-S4-t}, we introduce lTS4 and gTS4 and prove some basic propositions for lTS4 and gTS4. 

In Section \ref{equivalence-section-S4-t}, we define Kripke's calculus GS4, establish the equivalence among GS4, lTS4, and gTS4, and observe a comparison among proofs generated by lTS4, gTS4, and GS4.

In Section \ref{basic-theorems-S4-t}, we prove some basic theorems for lTS4 and gTS4. First, we show the classical-negation-elimination and classical-converse-negation-elimination theorems for lTS4 and gTS4. Second, we prove the cut-elimination theorems for lTS4 and gTS4, relying on key lemmas concerning the cut-free provabilities of lTS4, gTS4, and GS4. Finally, we obtain the subformula properties for lTS4 and gTS4 as a consequence of the cut-elimination theorems.

In Section \ref{other-systems-section}, we introduce Gentzen-style twist sequent calculi for other normal modal logics, including K, KT, and S5. Furthermore, we introduce a twist hyper-sequent calculus for S5. We also show the cut-elimination theorems and subformula properties for these calculi.

In Section \ref{conclusion-lTS4}, we conclude this study, offer some remarks on the potential applications of the proposed calculi to logic programming, and outline prospective future works.

\section{Twist sequent calculi for S4}
\label{twist-calculus-S4-t}

We construct {\em formulas} of normal modal logic S4 from countably many propositional variables by $\LAND$ (conjunction), $\LOR$ (disjunction), $\IMP$ (implication),  
$\neg$ (negation), $\BB$ (box), and  $\DIA$ (diamond). 
We use small letters $p, q, ...$ to denote propositional variables, Greek small letters $\al, \be,...$ to denote formulas, and Greek capital letters $\GA, \DE,...$ to represent finite (possibly empty) sets of formulas. 
For any set $A$ of symbols (i.e., alphabet), we use the notation $A^{\star}$ to represent the set of all words of finite length of the alphabet $A$. For any $\natural \in \{ \neg, \BB, \DIA \}^{\star}$, we use an expression $\natural \GA$ to denote the set $\{ \natural \ga \mid \ga \in \GA \}$. 
We use the symbol $\equiv$ to denote the equality of symbols. A {\em sequent} is an expression of the form \SEQ{\GA}{\DE}.
We use an expression $\al \Leftrightarrow \be$ to represent the abbreviation of the sequents $\SEQ{\al}{\be}$ and $\SEQ{\be}{\al}$.
We use an expression $L \vdash S$ to represent the fact that a sequent $S$ is provable in a sequent calculus $L$. 
We say that two sequent calculi $L_1$ and $L_2$ are {\em theorem-equivalent} if $\{S~|~ L_1 \vdash S\}$ $=$ $\{S~|~ L_2 \vdash S\}$. 
We say that a rule $R$ of inference is {\em admissible} in a sequent calculus $L$ if the following condition is satisfied: For any instance
{\small$
\frac{S_1 \cdots S_n}{S}
$}
of $R$, if $L \vdash S_i$ for all $i$, then $L \vdash S$. 
Furthermore, we say that $R$ is {\em derivable} in $L$ if there is a derivation from $S_1, \cdots, S_n $ to $S$ in $L$. We remark the fact that a rule $R$ of inference is admissible in a sequent calculus $L$ if and only if two sequent calculi $L$ and $L + R$ are theorem-equivalent. 
Since the logics discussed in this study are formulated as Gentzen-style sequent calculi, we will sometimes identify the logic with a Gentzen-style sequent calculus determined by it.

We introduce a Gentzen-style local twist sequent calculus lTS4 for S4.

\begin{df}[lTS4]
\label{GS4-3-sequent-calculus-definition}
The initial sequents of {\rm lTS4} are of the form:  
For any propositional variable $p$, 
$$
\SEQ{p}{p}
\quad\quad\quad
\SEQ{\neg p}{\neg p}
\quad\quad\quad
\SEQ{\neg p, p}{}
\quad\quad\quad
\SEQ{}{\neg p, p}.
$$

The structural inference rules of {\rm lTS4} are of the form:
$$
\infer[({\rm cut})]{\SEQ{\GA}{\DE}}{
   \SEQ{\GA}{\al}
    &
   \SEQ{\al,\GA}{\DE}
}
\quad
\infer[(\mbox{\rm we-left})]{\SEQ{\al,\GA}{\DE}}{
  \SEQ{\GA}{\DE}
}
\quad
\infer[(\mbox{\rm we-right}).]{\SEQ{\GA}{\DE, \al}}{
   \SEQ{\GA}{\DE}
}
$$

The non-twist logical inference rules of {\rm lTS4} are of the form: 
$$
\infer[(\LAND {\rm left})]{\SEQ{\al\LAND\be,\GA}{\DE}}{
   \SEQ{\al, \be, \GA}{\DE}
}
\quad
\infer[(\LAND {\rm right})]{\SEQ{\GA}{\DE, \al\LAND\be}}{
   \SEQ{\GA}{\DE, \al}
    &
   \SEQ{\GA}{\DE, \be}
}
$$
$$
\infer[(\LOR {\rm left})]{\SEQ{\al\LOR\be,\GA}{\DE}}{
      \SEQ{\al,\GA}{\DE}
       &
      \SEQ{\be, \GA}{\DE}
}
\quad
\infer[(\LOR {\rm right})]{\SEQ{\GA}{\DE, \al\LOR\be}}{
    \SEQ{\GA}{\DE, \al, \be}
} 
$$
$$
\infer[(\IMP {\rm left})]{\SEQ{\al \IMP \be,\GA}{\DE}}{
    \SEQ{\GA}{\DE, \al}
    &
    \SEQ{\be, \GA}{\DE}
}
\quad
\infer[(\IMP {\rm right})]{\SEQ{\GA}{\DE, \al \IMP \be}}{
   \SEQ{\al,\GA}{\DE, \be}
}
$$
$$
\infer[(\BB{\rm left})]{\SEQ{\BB\al, \GA}{\DE} }{
  \SEQ{\al, \GA}{\DE}
}
\quad
\infer[(\BB{\rm right})]{\SEQ{\BB\GA_1, \neg\DIA\GA_2}{\DIA\DE_1, \neg\BB\DE_2, \BB\al} }{
  \SEQ{\BB\GA_1, \neg\DIA\GA_2}{\DIA\DE_1, \neg\BB\DE_2, \al}
}
$$
$$
\infer[(\DIA{\rm left})]{\SEQ{\DIA\al, \BB\GA_1, \neg\DIA\GA_2}{\DIA\DE_1, \neg\BB\DE_2} }{
  \SEQ{\al, \BB\GA_1, \neg\DIA\GA_2}{\DIA\DE_1, \neg\BB\DE_2}
}
\quad
\infer[(\DIA{\rm right}).]{ \SEQ{\GA}{\DE, \DIA\al} }{
  \SEQ{\GA}{\DE, \al}
}
$$

The (local) twist logical inference rules (or twist rules for short) of {\rm lTS4} are of the form:
$$
\infer[(\neg\neg {\rm left}^t)]{\SEQ{\neg\neg\al, \GA}{\DE}}{
    \SEQ{\al, \GA}{\DE}
}
\quad
\infer[(\neg\neg {\rm right}^t)]{\SEQ{\GA}{\DE, \neg\neg\al}}{
    \SEQ{\GA}{\DE, \al}
}
$$
$$
\infer[(\mbox{$\neg\LAND$} {\rm left}^t)]{\SEQ{\neg (\al\LAND\be), \GA}{\DE}}{
      \SEQ{\GA}{\DE, \al}
       &
      \SEQ{\GA}{\DE, \be}
}
\quad
\infer[(\mbox{$\neg\LAND$} {\rm right}^t)]{\SEQ{\GA}{\DE, \neg(\al\LAND\be)}}{
    \SEQ{\al, \be, \GA}{\DE}
} 
$$
$$
\infer[(\mbox{$\neg\LOR$} {\rm left}^t)]{\SEQ{\neg (\al\LOR\be),\GA}{\DE}}{
   \SEQ{\GA}{\DE, \al, \be}
}
\quad
\infer[(\mbox{$\neg\LOR$} {\rm right}^t)]{\SEQ{\GA}{\DE, \neg (\al\LOR\be)}}{
   \SEQ{\al, \GA}{\DE}
    &
   \SEQ{\be, \GA}{\DE}
}
$$
$$
\infer[(\neg\IMP {\rm left}^t)]{\SEQ{\neg (\al\IMP\be),\GA}{\DE}}{
   \SEQ{\al, \GA}{\DE, \be}
}
\quad
\infer[(\neg\IMP {\rm right}^t)]{\SEQ{\GA}{\DE, \neg (\al\IMP\be)}}{
   \SEQ{\GA}{\DE, \al}
    &
   \SEQ{\be, \GA}{\DE}
}
$$
$$
\infer[(\neg\BB{\rm left}^t)]{ \SEQ{\neg\BB\al, \BB\GA_1, \neg\DIA\GA_2}{\DIA\DE_1, \neg\BB\DE_2}}{
  \SEQ{\BB\GA_1, \neg\DIA\GA_2}{\DIA\DE_1, \neg\BB\DE_2, \al}
}
\quad
\infer[(\neg\BB{\rm right}^t)]{\SEQ{\GA}{\DE, \neg\BB\al} }{
  \SEQ{\al, \GA}{\DE}
}
$$
$$
\infer[(\neg\DIA{\rm left}^t)]{ \SEQ{\neg\DIA\al, \GA}{\DE} }{
  \SEQ{\GA}{\DE, \al}
}
\quad
\infer[(\neg\DIA{\rm right}^t).]{\SEQ{\BB\GA_1, \neg\DIA\GA_2}{\DIA\DE_1, \neg\BB\DE_2, \neg\DIA\al}}{
  \SEQ{\al, \BB\GA_1, \neg\DIA\GA_2}{\DIA\DE_1, \neg\BB\DE_2}
}
$$
\end{df}


\begin{rmk}
\label{remark-lTS4}~
\begin{enumerate}
\item 
{\rm lTS4} has no standard logical inference rules for $\neg$ used in Gentzen's sequent calculus {\rm LK} {\rm \cite{GENTZEN}}:
$$
\infer[(\neg {\rm left})]{\SEQ{\neg\al, \GA}{\DE}}{
    \SEQ{\GA}{\DE, \al}
}
\quad
\infer[(\neg {\rm right}).]{\SEQ{\GA}{\DE, \neg\al}}{
    \SEQ{\al,\GA}{\DE}
}
$$
Instead, we use the twist logical inference rules in {\rm lTS4}. {\rm ($\IMP$left)} and {\rm ($\neg$right)} are internalized in the twist logical inference rules.

\item
The twist logical inference rules of {\rm lTS4} are constructed by integrating the (non-twist or standard) logical inference rules for $\LAND, \LOR, \IMP, \neg, \BB$, and $\DIA$ with the standard logical inference rules for $\neg$. 

\item
{\rm ($\neg\neg$left$^t$)} and {\rm ($\neg\neg$right$^t$)} are also constructed by integrating {\rm ($\neg$left)} with {\rm ($\neg$right)}. Thus, {\rm ($\neg\neg$left$^t$)} and {\rm ($\neg\neg$right$^t$)} are also said to be twist logical inference rules.

\item
Let {\rm lTS4$^{\star}$} be the system that is obtained from {\rm lTS4} by replacing {\rm ($\neg\BB$left$^t$)} and {\rm ($\neg\DIA$right$^t$)} with the simple twist rules of the form:
$$
\infer[(\neg\BB{\rm left}^{t\star})]{ \SEQ{\neg\BB\al, \BB\GA}{\DIA\DE}}{
  \SEQ{\BB\GA}{\DIA\DE, \al}
}
\quad
\infer[(\neg\DIA{\rm right}^{t\star}).]{\SEQ{\BB\GA}{\DIA\DE, \neg\DIA\al}}{
  \SEQ{\al, \BB\GA}{\DIA\DE}
}
$$
Then, the sequents of the form \SEQ{\neg\BB p}{\neg\BB p} and  \SEQ{\neg\DIA p}{\neg\DIA p} for any propositional variable $p$ cannot be proved in cut-free {\rm lTS4$^{\star}$}. Thus, we adopt {\rm ($\neg\BB$left$^t$)} and {\rm ($\neg\DIA$right$^t$)} in {\rm lTS4}.

\item 
{\rm ($\BB$right)} and {\rm ($\DIA$left)} in {\rm lTS4} are considered to be compatible with {\rm ($\neg\DIA$right$^t$)} and {\rm ($\neg\BB$left$^t$)}, respectively, in {\rm lTS4}. 
Actually, {\rm ($\neg\DIA$right$^t$)} and {\rm ($\neg\BB$left$^t$)} are constructed by integrating {\rm ($\BB$right)} and {\rm ($\DIA$left)} with {\rm ($\neg$left)} and {\rm ($\neg$right)}. {\rm ($\BB$right)} and {\rm ($\DIA$left)} are required for proving some basic properties. Thus, {\rm ($\BB$right)} and {\rm ($\DIA$left)} also cannot be replaced with the following simple rules:
$$
\infer[(\BB{\rm right}^k)]{\SEQ{\BB\GA}{\DIA\DE, \BB\al} }{
  \SEQ{\BB\GA}{\DIA\DE, \al}
}
\quad
\infer[(\DIA{\rm left}^k),]{\SEQ{\DIA\al, \BB\GA}{\DIA\DE} }{
  \SEQ{\al, \BB\GA}{\DIA\DE}
}
$$
which were used in Kripke's Gentzen-style sequent calculus (for {\rm S4}) originally introduced in {\rm \cite{KRIPKE-ZML1963}} (p. 91).


\item 
Let {\rm TCL} be the system that is obtained from {\rm lTS4} by deleting the logical inference rules concerning $\BB$ and $\DIA$ (i.e., {\rm TCL} is the $\{\BB, \DIA\}$-less fragment of {\rm lTS4}). Then, {\rm TCL} is theorem-equivalent to Gentzen's sequent calculus {\rm LK}  {\rm \cite{GENTZEN}} for propositional classical logic, and hence {\rm TCL} is a Gentzen-style twist sequent calculus for propositional classical logic. 
\end{enumerate}
\end{rmk}

Next, we introduce a Gentzen-style global twist sequent calculus gTS4 for S4.

\begin{df}[gTS4]
\label{GS4-3-sequent-calculus-definition-2}
{\rm gTS4} is obtained from {\rm lTS4} by replacing {\rm ($\BB$right)}, {\rm ($\DIA$left)}, {\rm ($\neg\BB$left$^t$)}, and {\rm ($\neg\DIA$left$^t$)} with the (global) twist logical inference rules of the form: 
$$
\infer[(\BB{\rm right}^T)]{\SEQ{\BB\GA_1, \neg\DIA\GA_2}{\DIA\DE_1, \neg\BB\DE_2, \BB\al} }{
  \SEQ{\BB\GA_1, \BB\DE_2}{\DIA\DE_1, \DIA\GA_2, \al}
}
\quad
\infer[(\DIA{\rm left}^T)]{\SEQ{\DIA\al, \BB\GA_1, \neg\DIA\GA_2}{\DIA\DE_1, \neg\BB\DE_2} }{
  \SEQ{\al, \BB\GA_1, \BB\DE_2}{\DIA\DE_1, \DIA\GA_2}
}
$$
$$
\infer[(\neg\BB{\rm left}^{T})]{ \SEQ{\neg\BB\al, \BB\GA_1, \neg\DIA\GA_2}{\DIA\DE_1, \neg\BB\DE_2}}{
  \SEQ{\BB\GA_1, \BB\DE_2}{\DIA\DE_1, \DIA\GA_2, \al}
}
\quad
\infer[(\neg\DIA{\rm right}^T).]{\SEQ{\BB\GA_1, \neg\DIA\GA_2}{\DIA\DE_1, \neg\BB\DE_2, \neg\DIA\al}}{
  \SEQ{\al, \BB\GA_1, \BB\DE_2}{\DIA\DE_1, \DIA\GA_2}
}
$$
\end{df}

\begin{rmk}
We now address a comparison between {\rm lTS4} and {\rm gTS4}. In a sense, {\rm lTS4} is a local calculus for handling $\neg$ and {\rm gTS4} is a global calculus for handling $\neg$. On the one hand, the twist logical inference rules for $\neg\BB$ and $\neg\DIA$ in {\rm lTS4} are applied only for the principal formulas $\neg\BB\al$ and $\neg\DIA\al$ of the twist rules. Namely, the occurrences of $\neg$ in the non-principal contexts of the lower sequents of the twist rules are retained in the upper sequents (i.e., $\neg$ is handled locally). On the other hand, the upper sequents of the twist rules for $\neg\BB$ and $\neg\DIA$ in {\rm gTS4} have no $\neg$. Namely, all the occurrences of $\neg$ in the contexts of the lower sequents of the twist rules are deleted in the upper sequents (i.e., $\neg$ is handled globally). Thus, we call {\rm lTS4} and {\rm gTS4} local and global twist calculi, respectively.  

\end{rmk}


\begin{prop}
\label{GS4-3-provable-sequents-proposition}
Let $L$ be {\rm lTS4} or {\rm gTS4}.
The following sequents are provable in cut-free $L$: For any formula $\al$,
\begin{enumerate}
\item
\label{GS4-3-initial-1} 
$\SEQ{\al}{\al}$,

\item
\label{GS4-3-initial-2} 
\SEQ{\al, \neg\al}{},

\item
\label{GS4-3-initial-3}
\SEQ{}{\al, \neg\al}.
\end{enumerate}
\end{prop}
\PROOF
We only prove the proposition for lTS4, because the proposition for gTS4 can be proved similarly. We now prove the statements \ref{GS4-3-initial-1} and \ref{GS4-3-initial-2} for lTS4. The statement \ref{GS4-3-initial-3} for lTS4 can be proved in a similar way as that for \ref{GS4-3-initial-2}. Thus, the proof of the statement \ref{GS4-3-initial-3} for lTS4 is omitted. 
\begin{enumerate}
\item
We prove the statement \ref{GS4-3-initial-1} by induction on $\al$. 
We distinguish the cases according to the form of $\al$ and show only the case $\al \equiv \neg\be$. 
In this case, we distinguish the cases according to the form of $\be$ and show some cases.  
\begin{enumerate}

\item
Case $\be \equiv \be_1\IMP\be_2$: 
We obtain the required proof: 
$$
\infer[(\neg\IMP{\rm left}^t).]{\SEQ{\neg(\be_1\IMP\be_2)}{\neg(\be_1\IMP\be_2)}}{
      \infer[(\neg\IMP{\rm right}^t)]{\SEQ{\be_1}{\neg(\be_1\IMP\be_2), \be_2}}{
             \infer[\mbox{(we-right)}]{\SEQ{\be_1}{\be_2, \be_1}}{
                  \infer*[Ind.~hyp.]{\SEQ{\be_1}{\be_1}}{
                  }
             }
             &
             \infer[\mbox{(we-left)}]{\SEQ{\be_2, \be_1}{\be_2}}{
                  \infer*[Ind.~hyp.]{\SEQ{\be_2}{\be_2}}{
                  }
             }
      }
}
$$


\item
Case $\be \equiv \BB\be_1$:
We can obtain the required proof: 
$$
\infer[(\neg\BB{\rm left}^t).]{\SEQ{\neg\BB\be_1}{\neg\BB\be_1}}{
      \infer[(\neg\BB{\rm right}^t)]{\SEQ{}{\neg\BB\be_1, \be_1}}{
            \infer*[Ind.~hyp.]{\SEQ{\be_1}{\be_1}}{
             }
       }
}
$$
We remark that we cannot prove this case using the simple rule ($\neg\BB$left$^{t\star}$) considered in Remark \ref{remark-lTS4}.
\end{enumerate}

\item
We prove the statement \ref{GS4-3-initial-2} by induction on $\al$.  
We distinguish the cases according to the form of $\al$ and show only the following cases. 
We have to prove some cases by using the statement 1. 
\begin{enumerate}

\item
Case $\al\equiv \be_1\IMP\be_2$: 
We obtain the required proof:
$$
\infer[(\neg\IMP{\rm left}^t).]{\SEQ{\be_1\IMP\be_2, \neg(\be_1\IMP\be_2)}{}}{
     \infer[(\IMP{\rm left})]{\SEQ{\be_1, \be_1\IMP\be_2}{\be_2}}{
           \infer[\mbox{(we-right)}]{\SEQ{\be_1}{\be_1, \be_2}}{
                 \infer*[Prop.~\ref{GS4-3-provable-sequents-proposition} (1)]{\SEQ{\be_1}{\be_1}}{
                 }
           }
           &
           \infer[\mbox{(we-left)}]{\SEQ{\be_1, \be_2}{\be_2}}{
                 \infer*[Prop.~\ref{GS4-3-provable-sequents-proposition} (1)]{\SEQ{\be_2}{\be_2}}{
                 }
           }
     }
}
$$


\item
Case $\al\equiv \BB\be$: 
We obtain the required proof:
$$
\infer[(\neg\BB{\rm left}^t).]{\SEQ{\BB\be, \neg\BB\be}{}}{
   \infer[(\BB{\rm left})]{\SEQ{\BB\be}{\be}}{
        \infer*[Prop.~\ref{GS4-3-provable-sequents-proposition} (1)]{\SEQ{\be}{\be}}{
         }
    }
}
$$
\end{enumerate}
\end{enumerate}
\QED


\section{Equivalence and comparison among calculi}
\label{equivalence-section-S4-t}

In this section, we define Kripke's Gentzen-style sequent calculus GS4 for S4 and show the theorem-equivalence among GS4, lTS4, and gTS4.

\begin{df}[GS4]
\label{GS4-definition} 
{\rm GS4} is obtained from {\rm lTS4} by replacing {\rm ($\BB$right)}, {\rm ($\DIA$left)}, all the twist logical inference rules, and the negated initial sequents of the form (\SEQ{\neg p}{\neg p}), (\SEQ{\neg p, p}{}), and (\SEQ{}{\neg p, p}) with the logical inference rules of the form: 
$$
\infer[(\neg {\rm left})]{\SEQ{\neg\al, \GA}{\DE}}{
    \SEQ{\GA}{\DE, \al}
}
\quad
\infer[(\neg {\rm right})]{\SEQ{\GA}{\DE, \neg\al}}{
    \SEQ{\al,\GA}{\DE}
}
\quad
\infer[(\BB{\rm right}^k)]{\SEQ{\BB\GA}{\DIA\DE, \BB\al} }{
  \SEQ{\BB\GA}{\DIA\DE, \al}
}
\quad
\infer[(\DIA{\rm left}^k).]{\SEQ{\DIA\al, \BB\GA}{\DIA\DE} }{
  \SEQ{\al, \BB\GA}{\DIA\DE}
}
$$
\end{df}

\begin{rmk}~
\begin{enumerate}
\item 
Strictly speaking, {\rm GS4} is regarded as a non-essential and small modification of Kripke's original Gentzen-style sequent calculus (for {\rm S4}) introduced in {\rm \cite{KRIPKE-ZML1963}} (p. 91) to deal with $\BB$ and $\DIA$ simultaneously. The original system by Kripke has the formula-based initial sequents of the form \SEQ{\al}{\al} for any formula $\al$ instead of the propositional-variable-based initial sequents. This original system was introduced by modifying Ohnishi and Matsumoto's Gentzen-style sequent calculus (for {\rm S4}) introduced in {\rm \cite{OM1957,OM1959}}. Some extensions and modifications of the system of this type have been recently introduced and studied by Grigoriev and Petrukhin in {\rm \cite{GP-LLP2019}} and by Kamide in {\rm \cite{KAMIDE-JLLI-2022}}.

\item 
The difference between Kripke's system (and its small modification {\rm GS4}) and Ohnishi and Matsumoto's system is the form of {\rm ($\BB$right$^k$)} and {\rm ($\DIA$left$^k$)}. Ohnishi and Matsumoto's system has no $\DIA\DE$ in {\rm ($\BB$right$^k$)} and $\BB\GA$ in {\rm ($\DIA$left$^k$)}. Using the rules of {\rm GS4}, we can show that the sequents of the form $\BB\al \Leftrightarrow \neg\DIA\neg\al$ and $\DIA\al \Leftrightarrow \neg\BB\neg\al$ for any formula $\al$ are provable in cut-free {\rm GS4}. These sequents cannot be proved in Ohnishi and Matsumoto's system. For more information on these characteristic rules, see {\rm \cite{KRIPKE-ZML1963,GP-LLP2019,KAMIDE-JLLI-2022}}.

\item 
The sequents of the form \SEQ{\al}{\al} for any formula $\al$ are provable in cut-free {\rm GS4}. This fact can be shown by induction on $\al$. Thus, we can take the sequents of the form \SEQ{\al}{\al} for any formula $\al$ as initial sequents of {\rm GS4}.

\item
The following rules are derivable in {\rm GS4} using {\rm (cut)}:
$$
\infer[(\neg {\rm left}^{-1})]{\SEQ{\al, \GA}{\DE}}{
  \SEQ{\GA}{\DE, \neg\al}
}
\quad
\infer[(\neg {\rm right}^{-1}).]{\SEQ{\GA}{\DE, \al}}{
  \SEQ{\neg\al, \GA}{\DE}
}
$$

\item 
The cut-elimination and Kripke-completeness theorems hold for Kripke's original system. Thus, the same  theorems also hold for {\rm GS4}. For more information on these theorems, see {\rm \cite{KRIPKE-ZML1963,GP-LLP2019}}. 
\end{enumerate}
\end{rmk}

\begin{thm}[Equivalence among {\rm lTS4}, {\rm gTS4}, and {\rm GS4}]
\label{equivalence-GS4-3-GS4}
Let $L$ be {\rm lTS4} or {\rm gTS4}.
The systems $L$ and {\rm GS4} are theorem-equivalent.
\end{thm}
\PROOF
We only prove the theorem for lTS4, because the proof of the theorem for gTS4 can be obtained similarly. 
Obviously, the negated initial sequents of {\rm lTS4} are provable in cut-free {\rm GS4}, and 
the negated logical inference rules of {\rm lTS4} are derivable in {\rm GS4}. 
For example, the derivability of ($\neg\BB$left$^t$) in {\rm GS4} is shown as follows. 
$$
\infer*[(\neg{\rm left}), (\neg{\rm right})]{\SEQ{\neg\BB\al, \BB\GA_1, \neg\DIA\GA_2}{\DIA\DE_1, \neg\BB\DE_2}}{
     \infer[(\BB{\rm right})]{\SEQ{\BB\GA_1, \BB\DE_2}{\DIA\DE_1, \DIA\GA_2, \BB\al}}{
           \infer*[(\neg{\rm left}^{-1}), (\neg{\rm right}^{-1})]{\SEQ{\BB\GA_1, \BB\DE_2}{\DIA\DE_1, \DIA\GA_2, \al}}{
                   \SEQ{\BB\GA_1, \neg\DIA\GA_2}{\DIA\DE_1, \neg\BB\DE_2, \al}
           }
     }
}
$$
where ($\neg$left$^{-1}$) and ($\neg$right$^{-1}$) are derivable in GS4 using (cut).  
Conversely, ($\neg$left) and ($\neg$right) in {\rm GS4} are derivable in {\rm lTS4} using (cut) by:
$$
\infer[({\rm cut})]{\SEQ{\neg\al, \GA}{\DE}}{
       \SEQ{\GA}{\DE, \al}
       &
       \infer*[{\rm Prop.}~\ref{GS4-3-provable-sequents-proposition}~(\ref{GS4-3-initial-2})]{\SEQ{\al,\neg\al}{}}{}
}
\quad
\infer[({\rm cut}).]{\SEQ{\GA}{\DE, \neg\al}}{
       \infer*[{\rm Prop.}~\ref{GS4-3-provable-sequents-proposition}~(\ref{GS4-3-initial-3})]{\SEQ{}{\al, \neg\al}}{}
       &
       \SEQ{\al, \GA}{\DE}
}
$$
Therefore, {\rm lTS4} and {\rm GS4} are theorem-equivalent. 
\QED

\begin{rmk}
The proofs generated by {\rm lTS4} and {\rm gTS4} are shorter than those of {\rm GS4}. Furthermore, both the proofs generated by {\rm lTS4} and {\rm gTS4} are composed of subformulas of the formulas included in the last sequent. 
If $\neg$ appears many times in a given provable sequent, then the generated proofs by {\rm lTS4} or {\rm gTS4} are quite shorter than those generated by {\rm GS4}. Thus, {\rm lTS4} and {\rm gTS4} are regarded as effective systems for proving negated modal formulas containing numerous negation connectives. We will illustrate a comparison among proofs generated by {\rm lTS4}, {\rm gTS4}, and {\rm GS4}.  
\end{rmk}

\begin{eg}
We consider the provable sequent \SEQ{\neg\neg\neg\DIA\neg p}{\neg\DIA\neg\neg\DIA\neg\neg\neg p} with a propositional variable $p$. 
The proofs of this sequent in {\rm lTS4}, {\rm gTS4}, and {\rm GS4} are addressed as follows. 
First, we show the short proof generated by {\rm lTS4} using the twist rules {\rm ($\neg\neg$left$^t$)}, {\rm ($\neg\DIA$left$^t$)}, and {\rm ($\neg\DIA$left$^t$)}
and the negated initial sequent \SEQ{\neg p}{\neg p}.  
$$
\infer[(\neg\neg{\rm left}^t).]{\SEQ{\neg\neg\neg\DIA\neg p}{\neg\DIA\neg\neg\DIA\neg\neg\neg p}}{
    \infer[(\neg\DIA {\rm right}^t)]{\SEQ{\neg\DIA\neg p}{\neg \DIA\neg\neg\DIA\neg\neg\neg p}}{
          \infer[(\neg\neg {\rm left}^t)]{\SEQ{\neg\neg\DIA\neg\neg\neg p, \neg\DIA\neg p}{}}{
                 \infer[(\DIA {\rm left})]{\SEQ{\DIA\neg\neg\neg p, \neg\DIA\neg p}{}}{
                        \infer[(\neg\neg {\rm left}^t)]{\SEQ{\neg\neg\neg p, \neg\DIA\neg p}{}}{
                               \infer[(\neg\DIA{\rm left}^t)]{\SEQ{\neg p, \neg\DIA\neg p}{}}{
                                    \SEQ{\neg p}{\neg p}
                               }
                         }
                  }
           }
    }
}
$$
Next, we show the short proof generated by {\rm gTS4} using the twist rules {\rm ($\neg\neg$left$^t$)}, {\rm ($\neg\DIA$right$^T$)}, and {\rm ($\DIA$left$^T$)} and the negated initial sequent \SEQ{\neg p}{\neg p}.  
$$
\infer[(\neg\neg{\rm left}^t).]{\SEQ{\neg\neg\neg\DIA\neg p}{\neg\DIA\neg\neg\DIA\neg\neg\neg p}}{
    \infer[(\neg\DIA {\rm right}^T)]{\SEQ{\neg\DIA\neg p}{\neg \DIA\neg\neg\DIA\neg\neg\neg p}}{
          \infer[(\neg\neg {\rm left}^t)]{\SEQ{\neg\neg\DIA\neg\neg\neg p}{\DIA\neg p}}{
                 \infer[(\DIA {\rm left}^T)]{\SEQ{\DIA\neg\neg\neg p}{\DIA\neg p}}{
                        \infer[(\neg\neg {\rm left}^t)]{\SEQ{\neg\neg\neg p}{\DIA\neg p}}{
                               \infer[(\DIA{\rm right})]{\SEQ{\neg p}{\DIA\neg p}}{
                                    \SEQ{\neg p}{\neg p}
                               }
                         }
                  }
           }
    }
}
$$
Finally, we show the usual (long) proof generated by {\rm GS4} using the standard logical inference rules {\rm ($\neg$left)} and {\rm ($\neg$right)}.
$$
\infer[(\neg {\rm right}).]{\SEQ{\neg\neg\neg\DIA\neg p}{\neg\DIA\neg\neg\DIA\neg\neg\neg p}}{
     \infer[(\neg {\rm left})]{\SEQ{\neg\neg\neg\DIA\neg p, \DIA\neg\neg\DIA\neg\neg\neg p}{}}{
           \infer[(\neg {\rm right})]{\SEQ{\DIA\neg\neg\DIA\neg\neg\neg p}{\neg\neg\DIA\neg p}}{
                 \infer[(\neg {\rm left})]{\SEQ{\neg\DIA\neg p, \DIA\neg\neg\DIA\neg\neg\neg p}{}}{
                     \infer[(\DIA {\rm left}^k)]{\SEQ{\DIA\neg\neg\DIA\neg\neg\neg p}{\DIA\neg p}}{
                          \infer[(\neg {\rm left})]{\SEQ{\neg\neg\DIA\neg\neg\neg p}{\DIA\neg p}}{
                               \infer[(\neg {\rm right})]{\SEQ{}{\DIA\neg p, \neg\DIA\neg\neg\neg p}}{
                                   \infer[(\DIA {\rm left}^k)]{\SEQ{\DIA\neg\neg\neg p}{\DIA\neg p}}{
                                      \infer[(\DIA {\rm right})]{\SEQ{\neg\neg\neg p}{\DIA\neg p}}{
                                         \infer[(\neg {\rm left})]{\SEQ{\neg\neg\neg p}{\neg p}}{
                                            \infer[(\neg {\rm right})]{\SEQ{}{\neg p, \neg\neg p}}{
                                               \infer[(\neg {\rm left})]{\SEQ{\neg p}{\neg p}}{
                                                  \infer[(\neg {\rm right})]{\SEQ{}{\neg p, p}}{
                                                     \SEQ{p}{p}
                                                  }
                                               }
                                            }
                                          }
                                      }
                                   }
                               }
                           }
                      }
                  }
            }
     }
}
$$
\end{eg}

\section{Cut-elimination and subformula property}
\label{basic-theorems-S4-t}

In this section, we prove some basic theorems for lTS4 and gTS4. 


\begin{thm}[Classical-negation-elimination for lTS4 and gTS4]
\label{GS4-3-admissibility-of-negation-rules}
Let $L$ be {\rm lTS4} or {\rm gTS4}.
The rules {\rm ($\neg$left)} and {\rm ($\neg$right)} are admissible in cut-free $L$. 
\end{thm}
\PROOF
We show only the admissibility of ($\neg$left), because the admissibility of ($\neg$right) can be shown similarly. 
We consider the proof of the form:  
$$
\infer[(\neg {\rm left}).]{\SEQ{\neg\al, \GA}{\DE}}{
   \infer*[P]{\SEQ{\GA}{\DE, \al}}{
   }
}
$$
Then, we prove the theorem by induction on $P$. 
We distinguish the cases according to the last inference of $P$ and show some cases. 
\begin{enumerate}
%
\item
Case ($\IMP$right):
The last inference of $P$ is of the form:
$$
\infer[(\IMP {\rm right})]{\SEQ{\GA}{\DE, \al_1 \IMP \al_2}}{
    \infer*[]{\SEQ{\al_1, \GA}{\DE, \al_2}}{
    }
}
$$
where $\al \equiv \al_1\IMP\al_2$.
We then obtain the required fact:
$$
\infer[(\neg\IMP{\rm left}^t).]{\SEQ{\neg(\al_1\IMP\al_2), \GA}{\DE}}{
    \infer*[]{\SEQ{\al_1, \GA}{\DE, \al_2}}{
    }
}
$$

\item
Case ($\neg\IMP$right$^t$): 
The last inference of $P$ is of the form:
$$
\infer[(\neg\IMP {\rm right}^t)]{\SEQ{\GA}{\DE, \neg (\al_1\IMP\al_2)}}{
   \infer*[]{\SEQ{\GA}{\DE, \al_1}}{
    }
    &
   \infer*[]{\SEQ{\al_2, \GA}{\DE}}{
    }
}
$$
where $\al \equiv \neg(\al_1 \IMP \al_2)$.
We then obtain the required fact:
$$
\infer[(\neg\neg{\rm left}^t).]{\SEQ{\neg\neg(\al_1\IMP\al_2), \GA}{\DE}}{
      \infer[(\I{\rm left})]{\SEQ{\al_1\IMP\al_2, \GA}{\DE}}{
            \infer*[]{\SEQ{\GA}{\DE, \al_1}}{
             }
             &
            \infer*[]{\SEQ{\al_2, \GA}{\DE}}{
             }
      }
}
$$

\item
Case ($\BB$right) for lTS4: 
The last inference of $P$ is of the form:
$$
\infer[(\BB{\rm right})]{\SEQ{\BB\GA_1, \neg\DIA\GA_2}{\DIA\DE_1, \neg\BB\DE_2, \BB\al_1} }{
   \infer*[]{\SEQ{\BB\GA_1, \neg\DIA\GA_2}{\DIA\DE_1, \neg\BB\DE_2, \al_1}}{
   }
}
$$
where $\SEQ{\GA}{\DE, \al}$ is $\SEQ{\BB\GA_1, \neg\DIA\GA_2}{\DIA\DE_1, \neg\BB\DE_2, \BB\al_1}$ and $\al \equiv \BB\al_1$.
We then obtain the required fact:
$$
\infer[(\neg\BB{\rm left}^t).]{\SEQ{\neg\BB\al_1, \BB\GA_1, \neg\DIA\GA_2}{\DIA\DE_1, \neg\BB\DE_2} }{
   \infer*[]{\SEQ{\BB\GA_1, \neg\DIA\GA_2}{\DIA\DE_1, \neg\BB\DE_2, \al_1}}{
   }
}
$$

\item
Case ($\neg\DIA$right$^t$) for lTS4: 
The last inference of $P$ is of the form:
$$
\infer[(\neg\DIA{\rm right}^t)]{\SEQ{\BB\GA_1, \neg\DIA\GA_2}{\DIA\DE_1, \neg\BB\DE_2, \neg\DIA\al_1} }{
   \infer*[]{\SEQ{\al_1, \BB\GA_1, \neg\DIA\GA_2}{\DIA\DE_1, \neg\BB\DE_2}}{
   }
}
$$
where $\SEQ{\GA}{\DE, \al}$ is $\SEQ{\BB\GA_1, \neg\DIA\GA_2}{\DIA\DE_1, \neg\BB\DE_2, \neg\DIA\al_1}$ and $\al \equiv \neg\DIA\al_1$.
We then obtain the required fact:
$$
\infer[(\neg\neg{\rm left}^t).]{\SEQ{\neg\neg\DIA\al_1, \BB\GA_1, \neg\DIA\GA_2}{\DIA\DE_1, \neg\BB\DE_2}}{
\infer[(\DIA{\rm left})]{\SEQ{\DIA\al_1, \BB\GA_1, \neg\DIA\GA_2}{\DIA\DE_1, \neg\BB\DE_2} }{
   \infer*[]{\SEQ{\al_1, \BB\GA_1, \neg\DIA\GA_2}{\DIA\DE_1, \neg\BB\DE_2}}{
   }
}
}
$$

\item
Case ($\BB$right$^T$) for gTS4: 
The last inference of $P$ is of the form:
$$
\infer[(\BB{\rm right}^T)]{\SEQ{\BB\GA_1, \neg\DIA\GA_2}{\DIA\DE_1, \neg\BB\DE_2, \BB\al_1} }{
   \infer*[]{\SEQ{\BB\GA_1, \BB\DE_2}{\DIA\DE_1, \DIA\GA_2, \al_1}}{
   }
}
$$
where $\SEQ{\GA}{\DE, \al}$ is $\SEQ{\BB\GA_1, \neg\DIA\GA_2}{\DIA\DE_1, \neg\BB\DE_2, \BB\al_1}$ and $\al \equiv \BB\al_1$.
We then obtain the required fact:
$$
\infer[(\neg\BB{\rm left}^T).]{\SEQ{\neg\BB\al_1, \BB\GA_1, \neg\DIA\GA_2}{\DIA\DE_1, \neg\BB\DE_2} }{
   \infer*[]{\SEQ{\BB\GA_1, \BB\DE_2}{\DIA\DE_1, \DIA\GA_2, \al_1}}{
   }
}
$$

\item
Case ($\neg\DIA$right$^T$) for gTS4: 
The last inference of $P$ is of the form:
$$
\infer[(\neg\DIA{\rm right}^T)]{\SEQ{\BB\GA_1, \neg\DIA\GA_2}{\DIA\DE_1, \neg\BB\DE_2, \neg\DIA\al_1} }{
   \infer*[]{\SEQ{\al_1, \BB\GA_1, \BB\DE_2}{\DIA\DE_1, \DIA\GA_2}}{
   }
}
$$
where $\SEQ{\GA}{\DE, \al}$ is $\SEQ{\BB\GA_1, \neg\DIA\GA_2}{\DIA\DE_1, \neg\BB\DE_2, \neg\DIA\al_1}$ and $\al \equiv \neg\DIA\al_1$.
We then obtain the required fact:
$$
\infer[(\neg\neg{\rm left}^t).]{\SEQ{\neg\neg\DIA\al_1, \BB\GA_1, \neg\DIA\GA_2}{\DIA\DE_1, \neg\BB\DE_2}}{
\infer[(\DIA{\rm left}^T)]{\SEQ{\DIA\al_1, \BB\GA_1, \neg\DIA\GA_2}{\DIA\DE_1, \neg\BB\DE_2} }{
   \infer*[]{\SEQ{\al_1, \BB\GA_1, \BB\DE_2}{\DIA\DE_1, \DIA\GA_2}}{
   }
}
}
$$
\end{enumerate}
\QED
\\

Next, we show the following theorem using Theorem \ref{GS4-3-admissibility-of-negation-rules}.

\begin{thm}[Classical-converse-negation-elimination for lTS4 and gTS4]
\label{GS4-3-admissibility-of-converse-negation-rules}
Let $L$ be {\rm lTS4} or {\rm gTS4}.
The following rules are admissible in cut-free $L$:
$$
\infer[(\neg {\rm left}^{-1})]{\SEQ{\al, \GA}{\DE}}{
  \SEQ{\GA}{\DE, \neg\al}
}
\quad
\infer[(\neg {\rm right}^{-1}).]{\SEQ{\GA}{\DE, \al}}{
  \SEQ{\neg\al, \GA}{\DE}
}
$$
\end{thm}
\PROOF
We only prove the theorem for lTS4. 
We show only the admissibility of ($\neg$left$^{-1}$). The admissibility of ($\neg$right$^{-1}$) can be shown similarly. 
We consider the proof of the form:  
$$
\infer[(\neg {\rm left}^{-1}).]{\SEQ{\al, \GA}{\DE}}{
    \infer*[P]{\SEQ{\GA}{\DE, \neg\al}}{
    }
}
$$
Then, we prove the theorem by induction on $P$. 
We distinguish the cases according to the last inference of $P$ and show some cases. 
\begin{enumerate}
\item 
Case ($\neg\neg$right$^t$):
The last inference of $P$ is of the form:
$$
\infer[(\neg\neg{\rm right}^t)]{\SEQ{\GA}{\DE, \neg\neg\al_1}}{
     \infer*[]{\SEQ{\GA}{\DE, \al_1}}{}
}
$$
where $\al \equiv \neg\al_1$. 
We then obtain the required fact: 
$$
\infer[(\neg{\rm left})]{\SEQ{\neg\al_1, \GA}{\DE}}{
     \infer*[]{\SEQ{\GA}{\DE, \al_1}}{}
}
$$
where ($\neg$left) is admissible in cut-free lTS4 by Theorem \ref{GS4-3-admissibility-of-negation-rules}.

\item
Case ($\neg\IMP$right$^t$):
The last inference of $P$ is of the form:
$$
\infer[(\neg\IMP {\rm right}^t)]{\SEQ{\GA}{\DE, \neg (\al_1\IMP\al_2)}}{
   \infer*[]{\SEQ{\GA}{\DE, \al_1}}{}
    &
   \infer*[]{\SEQ{\al_2, \GA}{\DE}}{}
}
$$
where $\al \equiv \al_1\IMP\al_2$.
We then obtain the required fact:
$$
\infer[(\IMP {\rm left}).]{\SEQ{\al_1\IMP\al_2, \GA}{\DE}}{
   \infer*[]{\SEQ{\GA}{\DE, \al_1}}{}
    &
   \infer*[]{\SEQ{\al_2, \GA}{\DE}}{}
}
$$

\item
Case ($\neg\DIA$right$^t$):
The last inference of $P$ is of the form:
$$
\infer[(\neg\DIA{\rm right}^t)]{\SEQ{\BB\GA_1, \neg\DIA\GA_2}{\DIA\DE_1, \neg\BB\DE_2, \neg\DIA\al_1}}{
   \infer*[]{\SEQ{\al_1, \BB\GA_1, \neg\DIA\GA_2}{\DIA\DE_1, \neg\BB\DE_2}}{}
}
$$
where $\SEQ{\GA}{\DE, \al}$ is $\SEQ{\BB\GA_1, \neg\DIA\GA_2}{\DIA\DE_1, \neg\BB\DE_2, \neg\DIA\al_1}$ and $\al \equiv \DIA\al_1$.
We then obtain the required fact:
$$
\infer[(\DIA{\rm left}).]{\SEQ{\DIA\al_1, \BB\GA_1, \neg\DIA\GA_2}{\DIA\DE_1, \neg\BB\DE_2}}{
   \infer*[]{\SEQ{\al_1, \BB\GA_1, \neg\DIA\GA_2}{\DIA\DE_1, \neg\BB\DE_2}}{}
}
$$
In this case, we note that ($\DIA$left) in lTS4 cannot be replaced with ($\DIA$left$^k$) in GS4. 
\end{enumerate}
\QED
\\

Next, we show the following lemma using Theorem \ref{GS4-3-admissibility-of-negation-rules}.


\begin{lm}
\label{GS4-3-cut-free-equivalence}
Let $L$ be {\rm lTS4} or {\rm gTS4}.
For any sequent $S$, if $S$ is provable in cut-free {\rm GS4}, then $S$ is provable in cut-free $L$.
\end{lm}
\PROOF
We only prove the theorem for {\rm lTS4}.
Suppose that a sequent \SEQ{\GA}{\DE} is provable in cut-free {\rm GS4}. 
Then, we show this lemma by induction on the cut-free proofs $P$ of \SEQ{\GA}{\DE}. 
We distinguish the cases according to the last inference of $P$ and show only the cases for ($\neg$left) and ($\neg$right). The proofs of these cases can be obtained using ($\neg$left) and ($\neg$right), which  are admissible in cut-free lTS4 by Theorem \ref{GS4-3-admissibility-of-negation-rules}.

\QED
\\

We show the following cut-elimination theorem using Lemma \ref{GS4-3-cut-free-equivalence}.

\begin{thm}[Cut-elimination for lTS4 and gTS4]
\label{cut-eli-GS4-3}
Let $L$ be {\rm lTS4} or {\rm gTS4}.
The rule {\rm (cut)} is admissible in cut-free $L$.
\end{thm}
\PROOF
We only prove the theorem for lTS4. 
Suppose that a sequent $S$ is provable in {\rm lTS4}.
Then, $S$ is provable in {\rm GS4} by Theorem \ref{equivalence-GS4-3-GS4}.
Thus, $S$ is provable in cut-free GS4 by the cut-elimination theorem for GS4. 
Thus, $S$ is provable in cut-free {\rm lTS4} by Lemma \ref{GS4-3-cut-free-equivalence}.
\QED


\begin{thm}[Subformula property for {\rm lTS4} and {\rm gTS4}]
\label{subformula-GS4-3}
Let $L$ be {\rm lTS4} or {\rm gTS4}.
The system $L$ has the subformula property. 
Namely, if a sequent $S$ is provable in $L$, then there is a proof $P$ of $S$ such that all formulas appear in $P$ are subformulas of some formula in $S$. 
\end{thm}
\PROOF
By a consequence of Theorem \ref{cut-eli-GS4-3}.
\QED

\begin{rmk}
{\rm lTS4} and {\rm gTS4} are conservative extensions of the Gentzen-style twist sequent calculus {\rm TCL} for propositional classical logic, which was considered in Remark \ref{remark-lTS4}. This fact is obtained by Theorem \ref{cut-eli-GS4-3}. The cut-elimination theorem and subformula property also hold for {\rm TCL}. 
\end{rmk}

\section{Twist sequent calculi for K, KT, and S5}
\label{other-systems-section}

First, we introduce Gentzen-style global twist sequent calculi gTK, gTKT, and gTS5 for K, KT, and S5, respectively.

\begin{df}[gTK, gTKT, and gTS5]
\label{gKT-definition}~
\begin{enumerate}
\item 
{\rm gTK} is obtained from {\rm gTS4} by replacing {\rm ($\BB$left)}, {\rm ($\BB$right$^T$)}, {\rm ($\DIA$left$^T$)}, {\rm ($\DIA$right)}, {\rm ($\neg\BB$left$^T$)}, {\rm ($\neg\BB$right$^t$)}, {\rm ($\neg\DIA$left$^t$)}, and {\rm ($\neg\DIA$right$^T$)} with the following global twist logical inference rules: 
$$
\infer[(\BB\mbox{\rm K-right}^T)]{\SEQ{\BB\GA_1, \neg\DIA\GA_2}{\DIA\DE_1, \neg\BB\DE_2, \BB\al}}{
  \SEQ{\GA_1, \DE_2}{\DE_1, \GA_2, \al}
}
$$
$$
\infer[(\DIA\mbox{\rm K-left}^T)]{\SEQ{\DIA\al, \BB\GA_1, \neg\DIA\GA_2}{\DIA\DE_1, \neg\BB\DE_2} }{
  \SEQ{\al, \GA_1, \DE_2}{\DE_1, \GA_2}
}
$$
$$
\infer[(\neg\BB\mbox{\rm K-left}^T)]{\SEQ{\neg\BB\al, \BB\GA_1, \neg\DIA\GA_2}{\DIA\DE_1, \neg\BB\DE_2} }{
  \SEQ{\GA_1, \DE_2}{\DE_1, \GA_2, \al}
}
$$
$$
\infer[(\neg\DIA\mbox{\rm K-right}^T).]{\SEQ{\BB\GA_1, \neg\DIA\GA_2}{\DIA\DE_1, \neg\BB\DE_2, \neg\DIA\al}}{
  \SEQ{\al, \GA_1, \DE_2}{\DE_1, \GA_2}
}
$$

\item
{\rm gTKT} is obtained from {\rm gTK} by adding {\rm ($\BB$left)}, {\rm ($\DIA$right)}, and the logical inference rules {\rm ($\neg\BB$right$^T$)} and {\rm ($\neg\DIA$left$^T$)}. 

\item
{\rm gTS5} is obtained from {\rm gTS4} by replacing {\rm ($\BB$right$^T$)}, {\rm ($\DIA$left$^T$)}, {\rm ($\neg\BB$left$^T$)}, {\rm ($\neg\DIA$right$^T$)} with the following global twist logical inference rules:
$$
\infer[(\BB\mbox{\rm S5-right}^T)]{\SEQ{\BB\GA_1, \neg\DIA\GA_2}{\BB\DE_1, \neg\DIA\DE_2, \DIA\LA_1, \neg\BB\LA_2, \BB\al}}{
  \SEQ{\BB\GA_1, \DIA\DE_2, \BB\LA_2}{\BB\DE_1, \DIA\LA_1, \DIA\GA_2, \al}
}
$$
$$
\infer[(\DIA\mbox{\rm S5-left}^T)]{\SEQ{\DIA\al, \BB\GA_1, \neg\DIA\GA_2, \DIA\SI_1, \neg\BB\SI_2}{\DIA\DE_1, \neg\BB\DE_2} }{
  \SEQ{\al, \BB\GA_1, \DIA\SI_1, \BB\DE_2}{\DIA\DE_1, \DIA\GA_2, \BB\SI_2}
}
$$
$$
\infer[(\neg\BB\mbox{\rm S5-left}^T)]{\SEQ{\neg\BB\al, \BB\GA_1, \neg\DIA\GA_2, \DIA\SI_1, \neg\BB\SI_2}{\DIA\DE_1, \neg\BB\DE_2} }{
  \SEQ{\BB\GA_1, \DIA\SI_1, \BB\DE_2}{\DIA\DE_1, \DIA\GA_2, \BB\SI_2, \al}
}
$$
$$
\infer[(\neg\DIA\mbox{\rm S5-right}^T).]{\SEQ{\BB\GA_1, \neg\DIA\GA_2}{\BB\DE_1, \neg\DIA\DE_2, \DIA\LA_1, \neg\BB\LA_2, \neg\DIA\al}}{
  \SEQ{\al, \BB\GA_1, \DIA\DE_2, \BB\LA_2}{\BB\DE_1, \DIA\LA_1, \DIA\GA_2}
}
$$
\end{enumerate}
\end{df}

\begin{rmk}
We can also consider the local-type twist sequent calculi {\rm lTKT} and {\rm lTS5}. However, we cannot consider the local-type twist sequent calculus {\rm lTK}. The Kripke-style non-twist sequent calculi for {\rm K}, {\rm KT}, and {\rm S5} were introduced and studied in {\rm \cite{KAMIDE-JLLI-2022}}. On the one hand, the cut-elimination theorems for the Gentzen-style twist sequent calculi {\rm lTS5} and {\rm gTS5} do not hold. A counter example sequent for this fact is \SEQ{p}{\BB\neg\BB\neg p} where $p$ is a propositional variable. This counterexample sequent was given by Takano in {\rm \cite{Takano-1992}} for the cut-elimination theorem for a standard Gentzen-style sequent calculus for {\rm S5}, introduced by Ohnishi and Matsumoto. On the other hand, we can show the cut-elimination theorem for a twist hypersequent calculus, {\rm HTS5}, for {\rm S5}. The cut-elimination theorem for {\rm HTS5} will be shown. In {\rm HTS5}, there is no distinction between local and global. 
For more information on hypersequent calculi for {\rm S5}, see e.g., {\rm\cite{P-1983,Avron-1996,Restall-2005,P-2008,Kurokawa-2013,Lahav-2013,BI-2015,GP-LLP2019,KAMIDE-JLLI-2022}} and the references therein. 
\end{rmk}

Next, we introduce a twist hypersequent calculus HTS5 for S5. 
We call an expression of the form $\SEQ{\GA_1}{\DE_1}~\mid~\cdots~\mid~\SEQ{\GA_n}{\DE_n}$ {\em hypersequent}. We define the hypersequent $\SEQ{\GA_1}{\DE_1}~\mid~\cdots~\mid~\SEQ{\GA_n}{\DE_n}$ as a finite multiset of sequents \SEQ{\GA_k}{\DE_k} $(1 \leq k \leq n)$. 
We use capital letters $H$, $G$, ... to represent hypersequents.

\begin{df}[HTS5]
\label{HTS5-definition}
The initial hypersequents of {\rm HTS5} are of the form: For any propositional variable $p$, 
$$
\SEQ{p}{p}
\quad\quad\quad
\SEQ{\neg p}{\neg p}
\quad\quad\quad
\SEQ{p, \neg p}{}
\quad\quad\quad
\SEQ{}{p, \neg p}.
$$

The structural inference rules of {\rm HTS5} are of the form:
$$
\infer[({\rm cut})]{\SEQ{\GA,\SI}{\DE,\PI}~\mid~H~\mid~G}{
   \SEQ{\GA}{\DE,\al}~\mid~H
    &
   \SEQ{\al,\SI}{\PI}~\mid~G
}
\quad
\infer[({\rm merge})]{\SEQ{\GA, \SI}{\DE, \PI}~\mid~H}{
      \SEQ{\GA}{\DE}~\mid~\SEQ{\SI}{\PI}~\mid~H
}
$$
$$
\infer[\mbox{\rm (in-we-left)}]{\SEQ{\al,\GA}{\DE}~\mid~H}{
  \SEQ{\GA}{\DE}~\mid~H
}
\quad
\infer[\mbox{\rm (in-we-right)}]{\SEQ{\GA}{\DE,\al}~\mid~H}{
   \SEQ{\GA}{\DE}~\mid~H
}
$$
$$
\infer[\mbox{\rm (ex-we-left)}]{\SEQ{\al}{}~\mid~H}{
   H
}
\quad
\infer[\mbox{\rm (ex-we-right).}]{\SEQ{}{\al}~\mid~H}{
   H
}
$$

The non-twist logical inference rules of {\rm HTS5} are of the form: 
$$
\infer[(\land {\rm left})]{\SEQ{\al\land\be,\GA}{\DE}~\mid~H}{
   \SEQ{\al, \be, \GA}{\DE}~\mid~H
}
\quad
\infer[(\land {\rm right})]{\SEQ{\GA}{\DE,\al\land\be}~\mid~H~\mid~G}{
   \SEQ{\GA}{\DE,\al}~\mid~H
    &
   \SEQ{\GA}{\DE,\be}~\mid~G
}
$$
$$
\infer[(\lor {\rm left})]{\SEQ{\al\lor\be,\GA}{\DE}~\mid~H~\mid~G}{
      \SEQ{\al,\GA}{\DE}~\mid~H
       &
      \SEQ{\be, \GA}{\DE}~\mid~G
}
\quad
\infer[(\lor {\rm right})]{\SEQ{\GA}{\DE, \al\lor\be}~\mid~H}{
    \SEQ{\GA}{\DE,\al, \be}~\mid~H
} 
$$
$$
\infer[(\I {\rm left})]{\SEQ{\al \I \be,\GA}{\DE}~\mid~H~\mid~G}{
    \SEQ{\GA}{\DE,\al}~\mid~H
    &
    \SEQ{\be, \GA}{\DE}~\mid~G
}
\quad
\infer[(\I {\rm right})]{\SEQ{\GA}{\DE, \al \I \be}~\mid~H}{
   \SEQ{\al,\GA}{\DE, \be}~\mid~H
}
$$
$$
\infer[(\BB{\rm left})]{\SEQ{\BB\al}{}~\mid~\SEQ{\GA}{\DE}~\mid~H}{
  \SEQ{\al, \GA}{\DE}~\mid~H
}
\quad
\infer[(\BB{\rm right})]{\SEQ{}{\BB\al}~\mid~H }{
  \SEQ{}{\al}~\mid~H
}
$$
$$
\infer[(\DIA{\rm left})]{\SEQ{\DIA\al}{}~\mid~H }{
  \SEQ{\al}{}~\mid~H
}
\quad
\infer[(\DIA{\rm right}).]{ \SEQ{\GA}{\DE}~\mid~\SEQ{}{\DIA\al}~\mid~H }{
  \SEQ{\GA}{\DE, \al}~\mid~H
}
$$

The twist logical inference rules of {\rm HTS5} are of the form: 
$$
\infer[(\neg\neg {\rm left})]{\SEQ{\neg\neg\al, \GA}{\DE}~\mid~H}{
    \SEQ{\al, \GA}{\DE}~\mid~H
}
\quad
\infer[(\neg\neg {\rm right})]{\SEQ{\GA}{\DE, \neg\neg\al}~\mid~H}{
    \SEQ{\GA}{\DE, \al}~\mid~H
}
$$
$$
\infer[(\mbox{$\neg\land$}{\rm left})]{\SEQ{\neg(\al\land\be), \GA}{\DE}~\mid~H~\mid~G}{
      \SEQ{\GA}{\DE, \al}~\mid~H
       &
      \SEQ{\GA}{\DE, \be}~\mid~G
}
\quad
\infer[(\mbox{$\neg\land$}{\rm right})]{\SEQ{\GA}{\DE, \neg(\al\land\be)}~\mid~H}{
    \SEQ{\al, \be, \GA}{\DE}~\mid~H
} 
$$
$$
\infer[(\mbox{$\neg\lor$}{\rm left})]{\SEQ{\neg (\al\lor\be),\GA}{\DE}~\mid~H}{
   \SEQ{\GA}{\DE, \al, \be}~\mid~H
}
\quad
\infer[(\mbox{$\neg\lor$}{\rm right})]{\SEQ{\GA}{\DE, \neg (\al\lor\be)}~\mid~H~\mid~G}{
   \SEQ{\al, \GA}{\DE}~\mid~H
    &
   \SEQ{\be, \GA}{\DE}~\mid~G
}
$$
$$
\infer[(\neg\I {\rm left})]{\SEQ{\neg (\al\I\be),\GA}{\DE}~\mid~H}{
   \SEQ{\al, \GA}{\DE, \be}~\mid~H
}
\quad
\infer[(\neg\I {\rm right})]{\SEQ{\GA}{\DE, \neg (\al\I\be)}~\mid~H~\mid~G}{
   \SEQ{\GA}{\DE, \al}~\mid~H
    &
   \SEQ{\be, \GA}{\DE}~\mid~G
}
$$
$$
\infer[(\neg\BB\mbox{\rm S5-left}^h)]{\SEQ{\neg\BB\al}{}~\mid~H }{
  \SEQ{}{\al}~\mid~H
}
\quad
\infer[(\neg\BB\mbox{\rm S5-right}^h)]{ \SEQ{\GA}{\DE}~\mid~\SEQ{}{\neg\BB\al}~\mid~H }{
  \SEQ{\al, \GA}{\DE}~\mid~H
}
$$
$$
\infer[(\neg\DIA\mbox{\rm S5-left}^h)]{\SEQ{\neg\DIA\al}{}~\mid~\SEQ{\GA}{\DE}~\mid~H}{
  \SEQ{\GA}{\DE, \al}~\mid~H
}
\quad
\infer[(\neg\DIA\mbox{\rm S5-right}^h).]{\SEQ{}{\neg\DIA\al}~\mid~H }{
  \SEQ{\al}{}~\mid~H
}
$$
\end{df}

\begin{thm}[Cut-elimination for {\rm gTK}, {\rm gTKT}, and {\rm HTS5}]
\label{cut-eli-gTK}
Let $L$ be {\rm gTK}, {\rm gTKT}, or {\rm HTS5}.
The rule {\rm (cut)} is admissible in cut-free $L$.
\end{thm}
\PROOF
Similar to the proof of Theorem \ref{cut-eli-GS4-3}. 
For the case of HTS5, we use a cut-free (non-twist) hypersequent calculus for S5, that includes the following standard logical inference rules for $\neg$:
$$
\infer[(\neg {\rm left})]{\SEQ{\neg\al, \GA}{\DE}~\mid~H}{
  \SEQ{\GA}{\DE, \al}~\mid~H
}
\quad
\infer[(\neg {\rm right}).]{\SEQ{\GA}{\DE, \neg\al}~\mid~H}{
  \SEQ{\al, \GA}{\DE}~\mid~H
}
$$ 
For more information on this standard hypersequent calculus, see \cite{Restall-2005,GP-LLP2019,KAMIDE-JLLI-2022}. 
\QED

\begin{thm}[Subformula property for {\rm gTK}, {\rm gTKT}, and {\rm HTS5}]
\label{subformula-gKT}
Let $L$ be {\rm gTK}, {\rm gTKT}, or {\rm HTS5}. 
The system $L$ has the subformula property. 
\end{thm}
\PROOF
By a consequence of Theorem \ref{cut-eli-gTK}.
\QED

\section{Concluding remarks}
\label{conclusion-lTS4}

In this study, we introduced and investigated the cut-free and analytic Gentzen-style local and global twist sequent calculi, lTS4 and gTS4, for the normal modal logic S4. In these calculi, negations are handled locally in lTS4 and globally in gTS4. Unlike standard calculi, lTS4 and gTS4 do not include standard logical inference rules for negation. Instead, they employ several twist logical inference rules, which serve as ``shortcut (or abbreviated)'' rules specifically for negated logical connectives. As a result, lTS4 and gTS4 can generate relatively short ``shortcut (or abbreviated)'' proofs for provable modal formulas containing numerous negation connectives.

We proved the cut-elimination theorems for lTS4 and gTS4 and obtained the subformula properties for them. Additionally, we observed that if a given provable modal formula contains numerous negation connectives, the lengths of the proofs generated by lTS4 and gTS4 are shorter than those generated by the standard Gentzen-style sequent calculus GS4. Thus, we have identified a method for generating short proofs for modal formulas containing numerous negation connectives. We also obtained similar results for the Gentzen-style twist sequent calculi, gTK and gTKT, for the normal modal logics K and KT, respectively. Additionally, we obtained a similar result for the twist hypersequent calculus, HTS5, for the normal modal logic S5.

On the one hand, as mentioned in Section \ref{other-systems-section}, we could construct the cut-free twist hypersequent calculus HTS5 for S5, in a similar way to those in \cite{GP-LLP2019,KAMIDE-JLLI-2022}. On the other hand, we have not yet considered other types of twist sequent calculi for S5 based on {\em tree-hypersequent calculi} studied by Poggiolesi and Lellmann \cite{Poggiolesi-2010,LP-2024}, {\em 2-sequent calculi} studied by Martini, Masini, and Zorzi \cite{MMZ-ACM-2021,MMZ-BSL-2023}, or {\em bisequent calculi} studied by Indrzejczak \cite{I-2019}. Additionally, in this study, we have not yet considered twist-style calculi in the usual sequent, hypersequent, tree-hypersequent, 2-sequent, or bisequent formats for non-normal modal logics. These issues are left as future work.

As mentioned in Section \ref{introduction-section}, reasoning about negative information or knowledge involving both negations and modalities holds significant importance in the field of philosophical logic. This type of reasoning is also crucial in computer science, particularly in logic programming and knowledge representation. Modal logic programming and knowledge representation involving modalities and negations have been extensively studied \cite{Pratt-1980,BG-JLP-1994,N-JANCL-2009,SE-AI-2016,Gelfond-2023}. In these areas, an effective proof system that can efficiently handle both modalities and negations simultaneously is required.

We believe that the proposed Gentzen-style twisted sequent calculi are useful for implementing a sequent calculus-based goal-directed logic programming language, known as a uniform proof-based abstract logic programming language, which was originally developed by Miller, Nadathur, Pfenning, and Scedrov \cite{MNPS}. In relation to this, abstract paraconsistent logic programming with uniform proof was studied by Kamide in \cite{KAMIDE2007}, where a uniform proof-theoretic foundation for that programming language, along with its applications, was proposed. Therefore, a promising future direction is to develop a uniform proof-theoretic abstract modal logic programming framework based on the proposed twisted sequent calculi, focusing on negations and modalities.

We also believe that shortcut (or abbreviated) reasoning, based on the proposed twist calculi, plays a crucial role in logic programming involving modalities and negations. This is because true negative information (or knowledge) in logic programming, represented by provable negated modal formulas containing modal operators and multiple negation connectives, often arises in real-world situations \cite{BG-JLP-1994,N-JANCL-2009,SE-AI-2016,Gelfond-2023}. In such cases, the proofs, which are often lengthy, are regarded as evidence. This evidence should be concise and ideally represented by short and compact shortcut (or abbreviated) proofs. In this context, short proofs are valuable and necessary for explaining evidence concisely.

\

\noindent {\bf Acknowledgments.} 
I would like to thank the anonymous referees for their valuable comments and suggestions. 
This research was supported by JSPS KAKENHI Grant Number 23K10990.

\nocite{*}
\bibliographystyle{eptcs}
\bibliography{generic}
\end{document}